%% file: paper.tex
\documentclass[reprint,amsmath,amssymb,aps,superscriptaddress]{revtex4-1} 
\usepackage{graphicx}
\usepackage{physics}
\usepackage{amsthm}
\usepackage{amssymb}
\usepackage{import}
\usepackage[utf8]{inputenc}
\usepackage[colorlinks]{hyperref}

\begin{document}
    \title{Qubit Motion as a Microscopic Model for the Dynamical Casimir Effect}
    \author{A. Agustí}
    \affiliation{Instituto de Física Fundamental, CSIC, Serrano 113-bis 28006 Madrid, Spain}
    \author{L. García-Álvarez}
    \affiliation{Department of Microtechnology and Nanoscience (MC2), Chalmers University of Technology, SE-412 96 Göteborg, Sweden}
    \author{E. Solano}
    \affiliation{International Center of Quantum Artificial Intelligence for Science and Technology (QuArtist) \\ and Physics Department, Shanghai University, 200444 Shanghai, China}\affiliation{Department of Physical Chemistry, University of the Basque Country UPV/EHU, Apartado 644, 48080 Bilbao, Spain}\affiliation{IKERBASQUE, Basque Foundation for Science, Plaza Euskadi 5, 48009 Bilbao, Spain}\affiliation{Kipu Quantum, Munich, Germany}
    \author{C. Sabín}
    \affiliation{Instituto de Física Fundamental, CSIC, Serrano 113-bis 28006 Madrid, Spain}
    \email{soyandres2@gmail.com}
    \begin{abstract}
        The generation of photons from the vacuum by means of the movement of a mirror is known as the dynamical Casimir effect (DCE). In general, this phenomenon is effectively described by a field with time-dependent boundary conditions. Alternatively, we introduce a microscopic model of the DCE capable of capturing the essential features of the effect with no time-dependent boundary conditions. Besides the field, such a model comprises a new subsystem representing the mirror's internal structure. In this work, we study one of the most straightforward mirror systems: a qubit moving in a cavity and coupled to one of the bosonic modes. We find that under certain conditions on the qubit's movement that do not depend on its physical properties, a large number of photons may be generated without changing the qubit state, as should be expected for a microscopic model of the mirror.
    \end{abstract}
    \maketitle

        In his seminal paper of 1970, G. Moore \cite{Moore_1970} discovered that relativistic movement of perfectly conducting mirrors could produce radiation even if the state of the electromagnetic field before the mirrors' movement were the vacuum. In the next years, the phenomenon was subsequently studied \cite{dewitt1975, fulling_davies1976, unknown1977} until the name \textit{Dynamical Casimir Effect} (DCE) was coined \cite{yablonovitch1989}, joining the broad family of quantum vacuum fluctuation effects that includes, among others, the Lamb shift \cite{PhysRev.72.241}, the Casimir-Polder effect \cite{PhysRev.73.360, casimir1948b} and the Unruh \cite{fulling1973, davies1975, unruh1976} and Hawking's radiations \cite{HAWKING1974}, to name a few \cite{nation_johansson2012}. For a long time, the realization of the DCE and most of these effects remained out of reach due to the experimental requirements to access the quantum and relativistic parameter regime needed for a measurable signal \cite{kim_brownell2006}. This hurdle was overcome this century with the advent of circuit quantum electrodynamics, as it allows experiments in the strong light-matter coupling regime \cite{Wallraff2004} and lead to a number of proposals for experimental observations of the DCE \cite{johansson_johansson2009,johansson_johansson2010}. In 2011, Wilson \textit{et al.} \cite{wilson_johansson2011} carried out an experiment in which the relativistically moving mirror was reproduced using a modulated magnetic flux threading a superconducting quantum interference device, leading to a time-dependent boundary condition in a microwave waveguide and a non-classical DCE photon production \cite{johansson_johansson2013}. The DCE was also observed in a Josephson metamaterial capable of modulating its refractive index \cite{lahteenmaki_paraoanu2013}, leading to an equivalent setting in which the effective length of a cavity changes over time.

        All the works cited so far model the moving mirrors as time-dependent boundary conditions. However, this boundary condition is an effective description that reproduces the effects of a more complex system, disregarding its microscopic features. In this work, we are interested in formulating a model that captures some of those microscopical features and reproduces the DCE with no time-dependent boundary conditions. The earliest application of this idea of an underlying \textit{microscopic model} dates back to the Ewald-Oseen extinction theorem \cite{ewald_hollingsworth1970, oseen1915, fearn_james1996,born-and-wolf}. According to the theorem, transmission and reflection of a plane wave at the interface between dielectric media can be understood as the collective response of the media's dipoles. This approach was applied by de Souza \textit{et al.} \cite{souza_impens2018} to model moving mirrors in a quantum field theory leading to the DCE.

\begin{figure}                   
            \includegraphics[scale=1]{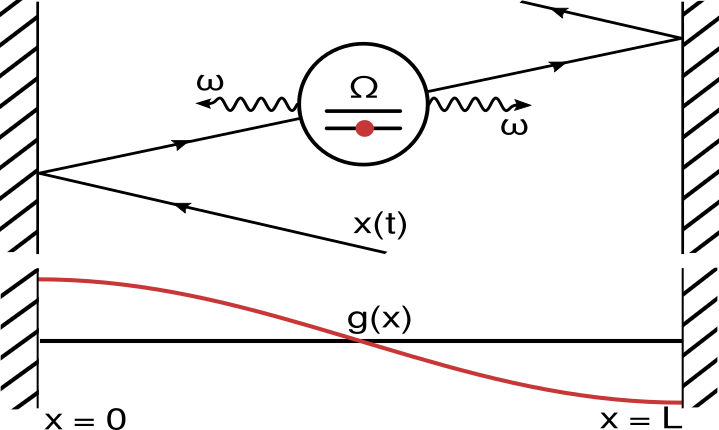}
            \caption{(Upper panel) Diagram of the system proposed as a microscopic model for the DCE. A qubit with frequency $\Omega$ and in its ground state moves back and forth inside a cavity of length $L$, producing photons in the fundamental mode of frequency $\omega$ if the qubit speed is close to the speed of light in the medium while staying in its ground state. Said photons are not produced by time-dependent boundary conditions, the walls (dashed dark blocks) are static. (Lower panel) Qubit-mode coupling $g$ as a function of qubit position for the fundamental mode, as defined in Eqs. (\ref{eq:H}) and (\ref{eq:coupling}). Then, the time dependent coupling is due to its composition with the trajectory and a small abuse in notation $g(t) = g[x(t)]$.}
            \label{fig:esquema}
        \end{figure}

        Following those steps, the goal of this work is to find a microscopic model for a moving mirror that reproduces the DCE and employs the ever-growing tool-set of present-day quantum technologies. We study the most straightforward system that may accommodate this phenomenon: a discrete mirror corresponding to a qubit moving in a cavity and interacting with one of its bosonic modes, as depicted in Figure \ref{fig:esquema}. The interaction between a qubit and a cavity has been studied extensively in the literature, especially in the case of a static qubit with the well-known Rabi model \cite{kellogg_rabi1936, braak2011}, and in the rotating wave approximation (RWA) regime with the Jaynes-Cummings (JC) model \cite{jcmodel}. Prior work has addressed the case of a moving qubit \cite{felicetti_sabin2015}, and in particular, the photon and qubit excitation due to the so-called cavity-enhanced Unruh effect \cite{scully_kocharovsky2003, PhysRevLett.93.129301, PhysRevLett.93.129302, PhysRevA.74.023807}. However, the parameter regime found to reproduce the DCE, where both rotating and counter-rotating terms are relevant, has not been explored before, to the authors' knowledge. In this article, we show that the qubit motion generates photons without changing its internal state for speeds close to the speed of light in the medium. This behavior provides a microscopic model for the DCE.

        The structure of this article is as follows: First, we recall the conditions of a microscopic model for the DCE and analyze the qubit-cavity system that fulfills the requirements. Second, we numerically explore the system's parameter space, finding both the JC model and the cavity-Unruh regime, as well as the novel \textit{microscopic} DCE regime. Then, we present a perturbative analytical justification for this latter regime, leading to the characterization of the microscopic DCE as a second-order perturbation theory effect. Finally, we conclude with a summary of the main results and our remarks about possible experimental implementations. 

        \paragraph*{Microscopic DCE with a qubit-cavity system.---}
        The DCE consists in the generation of photons through the relativistic movement of a perfect mirror. To identify a system as a microscopic model of the time-dependent boundary producing the DCE, we first need to consider what properties are characteristic of the DCE. Firstly, the movement of the mirror triggers the generation of photons, and there is no generation if the mirror is static. Secondly, a perfect mirror does not take energy from the field, if anything the former will return any energy to the latter in a short amount of time. Thirdly, the predicted evolution of the DCE field is unitary \cite{Moore_1970}, and because of this, the global unitary must factor into two unitaries for the field and the mirror, producing no entanglement. Lastly, we find that, in all the DCE settings explored so far, few if not none make assumptions on the internal structure of the mirror, that is, its static Hamiltonian's spectrum. Regardless of how it compares to the field eigenenergies all settings expect the mirror to behave as an inert boundary condition. In order to propose a new microscopic system that reproduces all these characteristics we require said model to follow three requirements: 1) Its movement must trigger the generation of photons, 2) It must stay in its ground state, so that it does not take energy from the field and does not get entangled with it, and 3) Its static Hamiltonian's spectrum must play no role in the effect. In previous work \cite{souza_impens2018}, an oscillating atom moving non-relativistically in free space was proposed as a microscopic model for the DCE. Such a model fulfills our first two requirements, but not the third one. In the case of Ref. \cite{souza_impens2018}, the atom's energy gap must be large compared to the photon and movement frequencies. Other models treat the mirror as a system rather than a boundary condition but do not meet the above requirements \cite{macri_ridolfo2018, stefano_settineri2019,lo_fong2020,dodonov_militello2016}. These are valuable generalizations of the DCE to new regimes, although they do not fit our definition of a microscopic model.

        The discrete mirror that we find to follow these requirements is a point-like electric dipole qubit coupled to one bosonic field mode. It is described by the Hamiltonian
        \begin{align}
            H_{\text{total}} &= H_0 + H_{\text{int}} \nonumber \\
            H_0 &= \frac{\Omega}{2} \sigma_z 
            + \omega a^\dagger a \nonumber \\
            H_{\text{int}} &= g(t)\sigma_x(a^\dagger + a),
            \label{eq:H}
        \end{align}
        with $\Omega$ and $\omega$ the qubit and mode frequencies, $\sigma_x, \sigma_z$ the first and third Pauli matrices, $a^\dagger$ and $a$ the creation and annihilation operators for the bosonic mode, and $g$ the time-dependent coupling that varies due to the classical motion of the discrete mirror qubit. The coupling to the fundamental bosonic mode of a cavity with perfectly conducting and static edges takes the form
        \begin{align}
            \label{eq:coupling}
            g(t) &= g_0 \cos[kx(t)] \nonumber \\
            k &= \pi/L, 
        \end{align}
        where $L$ is the length of the cavity and $x$ the trajectory of the discrete mirror qubit. The coupling's cosine dependence with the qubit's position $x$ results from the cavity's fundamental mode field amplitude at different locations. Following the tradition of DCE models, we will consider the movement of the mirror $x(t)$ as externally prescribed and not a dynamical variable. This way the system must be regarded as open and driven from the outside so that energy conservation does not inhibit photon production. In the context of superconducting circuits, the coupling intensity, $g_0$, is typically one or two orders of magnitude smaller than the circuit frequencies \cite{Wallraff2004,srinivasan_hoffman2011}. Also, coupling modulation usually lies in utilizing superconducting quantum interference devices, instead of physically moving circuit components. Following that approach, the Hamiltonian of Eq.~(\ref{eq:H}) can be engineered using tunable-coupling qubits \cite{srinivasan_hoffman2011}. Alternatively, a promising candidate consists of film bulk acoustic resonators, which could behave as an actual moving discrete mirror \cite{wang_blencowe2019, oconnell_hofheinz2010}.

        Given the linearity of the Schrödinger equation, we expect a coupling modulated with a cosine shape of constant frequency to be the most appropriate for analytical calculations. Such cosine coupling modulation is produced by qubit trajectories $x = vt$ with constant velocity $v$, for which the coupling oscillates with a driving frequency $\omega_d = \pi v / L$. To bound the qubit trajectory to the cavity while keeping this pure cosine coupling, we invert the direction of the qubit's (otherwise constant) velocity every time it reaches the cavity edges. In order to relate the velocity with the rest of parameters of the system, we will pay attention to it in adimensional units $v/c_n = \omega_d/\omega$, with $c_n$ the speed of light in the medium. Moreover, we remind the reader that typical values of $c_n$ in superconducting circuit setups are $c_n/c \approx 0.4$ \cite{wilson_johansson2011}.

        \begin{figure}
            \includegraphics[scale=.275]{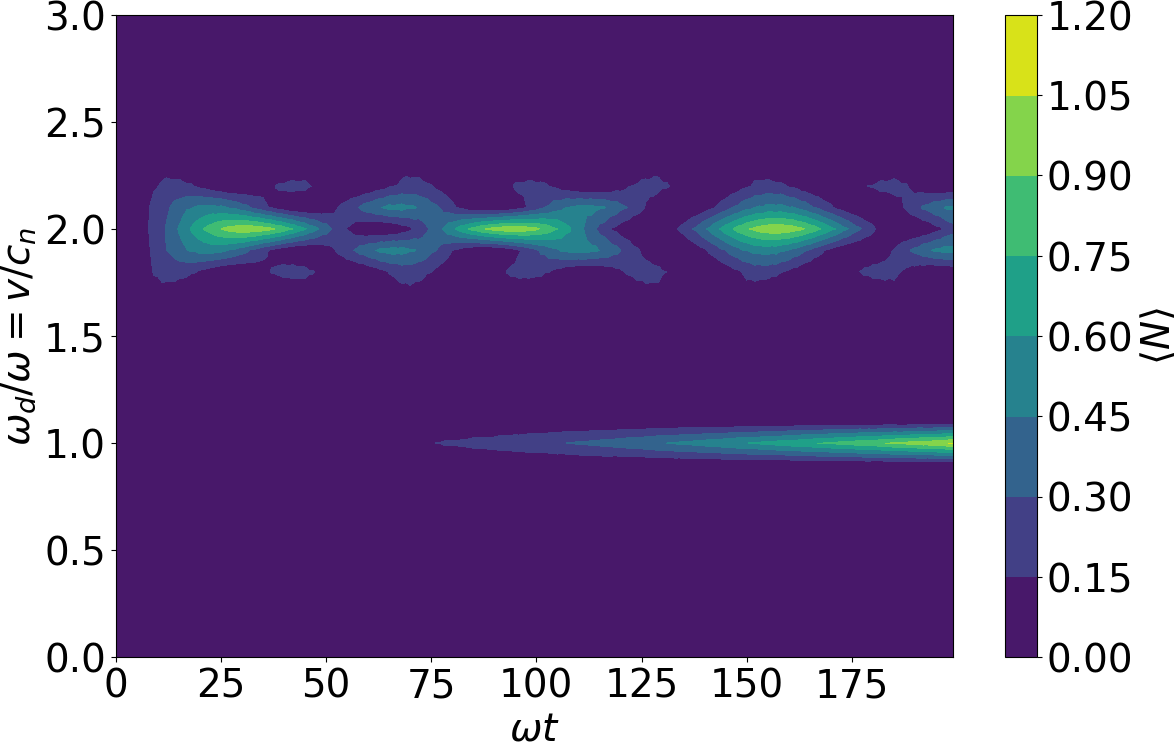}
            \caption{Number of photons $\expval{N}$ as a function of time $t$ in units of the mode frequency $\omega$, and driving frequency $\omega_d$ in units of the same mode frequency $\omega$, or equivalently, qubit velocity $v$ in units of the speed of light in the medium $c_n$. The driving frequency was produced by a qubit moving back and forth within the cavity, with constant velocity $v = L/\pi\omega_d$ in one direction, and $-v$ after bouncing in the opposite direction. The qubit frequency is given by $\Omega = \omega$, and the coupling intensity is $g_0 = 0.1\omega$.}
            \label{x-is-vt-N}
        \end{figure}
        \begin{figure}
            \includegraphics[scale=.275]{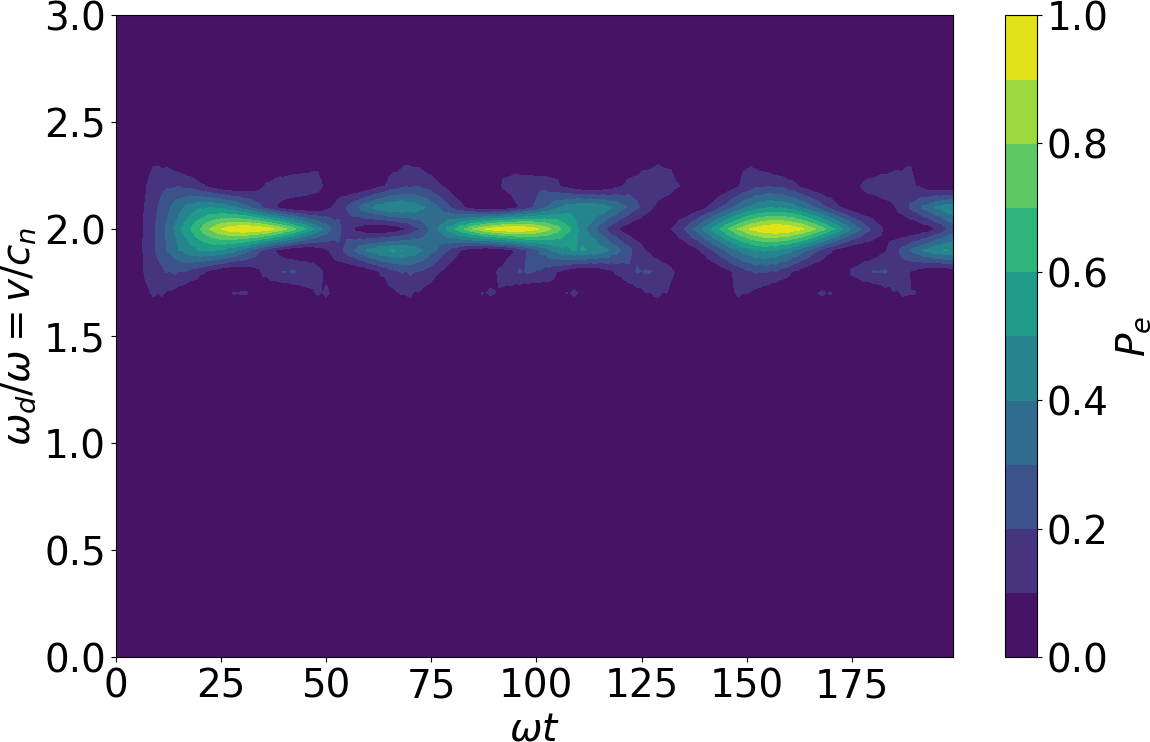}
            \caption{Qubits excited state population $P_e$, that is $\expval{\sigma^+\sigma^-}$, as a function of time $t$ in units of the mode frequency $\omega$, and driving frequency $\omega_d$ in units of the same mode frequency $\omega$ in the same conditions than Figure~\ref{x-is-vt-N}.}
            \label{x-is-vt-Sz}
        \end{figure}

        \paragraph*{Parameter space and physical regimes of the system.---}
        We numerically explore the parameter space of the Hamiltonian of Eq.~(\ref{eq:H}) describing the qubit-cavity system and identify three physical regimes. Namely, the JC regime, the cavity-enhanced Unruh regime, and the microscopic DCE regime.
        For the sake of simplicity, we consider the resonant case with equal mode and qubit frequencies, that is $\omega = \Omega$ respectively,  in the following simulations. We show the average cavity photon number $\expval{N}$ and the mean value of the qubit excited-state population $P_e = \expval{\sigma^{+}\sigma^{-}}$ as a function of time and qubit velocity in Figures \ref{x-is-vt-N} and \ref{x-is-vt-Sz}, respectively. The qubit-cavity system evolves from its ground state, for which $\expval{N}=0$ and $\expval{\sigma^+\sigma^-} = 0$.
        
        Firstly, the case of a static qubit corresponds to the JC model. Indeed, when the qubit velocity is zero ($\omega_d =0$), the coupling to the cavity is constant, $g(t) =g_0$ and for a coupling intensity $g_0 = 0.1\omega$, the rotating wave approximation (RWA) and the JC model hold. The counter-rotating terms $a^\dagger \sigma^+$ and $a \sigma^-$ produce no dynamics given their fast phase in the interaction picture. The rotating terms $a^\dagger \sigma^-$ and $a \sigma^+$ that compose the JC model do not generate dynamics either, given the initial ground state of the system, which remains unchanged as shown in Figures \ref{x-is-vt-N} and \ref{x-is-vt-Sz}.

        Alternatively, the \textit{anti-}RWA holds when the driving frequency is the sum of the qubit and bosonic mode frequencies $\omega_d = \omega + \Omega$. For the resonant case $\omega = \Omega$ studied, it corresponds to a qubit velocity twice the speed of light in the medium. In this case, we neglect the now fast-oscillating rotating terms, leaving the constant counter-rotating ones to generate the cavity-enhanced Unruh effect. The qubit is not undergoing a uniform acceleration motion as in the canonical Unruh effect \cite{fulling1973,davies1975,unruh1976}. Therefore, the radiation is not thermal, as it happens with more general trajectories \cite{obadia_milgrom2007,doukas_lin2013}. The common feature to these phenomena is dominant counter-rotating terms in the dynamics, as Scully \textit{et al.} noticed and exploited when developing the cavity-enhanced Unruh effect \cite{scully_kocharovsky2003}. 
        This relativistic effect has been studied previously for a qubit-cavity system \cite{felicetti_sabin2015}. There, the ground state of the system evolves towards the qubit excited state and one cavity photon for this qubit velocity, $v = 2 c_n$, in a Rabi-like oscillation. Consequently, both the photon number $\expval{N}$ (Figure \ref{x-is-vt-N}) and the qubit excited-state population $P_e$ (Figure \ref{x-is-vt-Sz}) increase up to one for $\omega_d/\omega = 2$ in our numerical simulations. 
        
        The third regime corresponds to the proposed \textit{microscopic} Dynamic Casimir effect. We observe a photon generation in the cavity (Figure \ref{x-is-vt-N}) while the qubit remains in its ground state (Figure \ref{x-is-vt-Sz}) for $\Omega=\omega=\omega_d$. This non-oscillatory monotonic photon production without qubit excitation occurs when the qubit moves at the speed of light in the medium, $\omega_d/\omega = v/c_n =1$. 
        As a final note on the numerical results, if the system is evolved further in time, the increasing number of photons requires a bigger subspace of the Hilbert space to be considered in the simulations.
        See Appendix~\ref{app:numerics} for proof that said subspace was large enough.

        \paragraph*{Perturbative analysis of the microscopic DCE model.---}
        A perturbative approach can explain the \textit{microscopic} DCE regime parameters, as it happens with the Unruh effect or the more widely known JC model. In order to make clear our discussion let us briefly fix some notation. Let the state be written as a power series on the coupling:
        \begin{align}
            \psi(t; g_0) = 
            \psi^{(0)} + 
            \psi^{(1)}(t)g_0 + 
            \frac{1}{2}\psi^{(2)}(t)g_0^2 + ...
            \label{series}.
        \end{align} 
        Each of the terms $\psi^{(n)}(t)$ corresponds to the $n$-th partial derivative of the state with respect to $g_0$, evaluated at $g_0 = 0$,
        \begin{align}
            \psi^{(n)}(t) = \partial^n_{g_0}\psi(t; g_0 = 0).
        \end{align}
        Although the functional dependence of $\psi(t; g_0)$ is hardly ever known, time-dependent perturbation theory enables us to compute its derivatives recursively as follows
        \begin{align}
            \psi^{(n+1)}(t) &= \int_0^t dt' \frac{H^{\mathcal{I}}_{\text{int}}(t')}{g_0} \psi^{(n)}(t')
            \label{eq:iteration_step}
        \end{align}
        where $H^{\mathcal{I}}_{\text{int}}$ is the interaction Hamiltonian $H_\text{int}$ in the interaction picture with respect to the static Hamiltonian term $H_0$.

        Following this method, we approximate the state of the system considering the Hamiltonian of Eq.~(\ref{eq:H}) and the initial ground state when computing the corrections from Eq.~(\ref{eq:iteration_step}). We remind the reader that for the static qubit case, the JC model is a good approximation because the rotating terms $a^\dagger\sigma^-$ and $a\sigma^+$, that the JC model shares with Hamiltonian Eq. (\ref{eq:H}), do not oscillate over time. Those terms produce perturbative corrections linear in time when integrated in Eq. (\ref{eq:iteration_step}). We define a \textit{resonance} as those integrals that result in linear contributions. Alternatively, a superluminal qubit's speed of $|v|=(1 + \Omega/\omega)c_n$ makes the counter-rotating terms resonate and the \textit{anti-}RWA becomes a good approximation, leading to the cavity-enhanced Unruh effect.

        In our main case of study, the \textit{microscopic} DCE, the qubit moves at a relativistic velocity $v \approx c_n$, leading to a driving of $\omega_d \approx \omega$. Then, both rotating and counter-rotating terms oscillate and cannot dominate the dynamics. Therefore, it is not straightforward to find a simplified Hamiltonian that behaves as the complete Hamiltonian of Eq. (\ref{eq:H}). If we consider a qubit moving at exactly the speed of light in the medium $v = c_n$ and apply twice Eq. (\ref{eq:iteration_step}) we find the second order correction
        \begin{align}
            & \nonumber
            \psi^{(2)}(t) = 
            \int_0^tdt'
            \int_0^{t'}dt''
            \frac{H^{\mathcal{I}}_{\text{int}}(t')}{g_0}\frac{H^{\mathcal{I}}_{\text{int}}(t'')}{g_0}
            \ket{g,0} = \\
            &
            \int_0^tdt'
            \int_0^{t'}dt''
            e^{-i\omega t'}\sigma^-a^\dagger
            e^{-i\omega t''}e^{i2\omega t''}\sigma^{+}a^\dagger
            \ket{g,0} + O(t^0),
            \label{eq:second_order_int}
        \end{align}
        where $\ket{g}$ is the qubit ground state and $\ket{n}$ is the $n$-photon state in the cavity fundamental mode. With $O(t^0)$, we indicate that we neglect any term bounded by a constant for long enough times. In our case, we disregard constant terms and exponentials with imaginary arguments. The oscillatory coupling leads to the $e^{-i\omega t}$ terms, and the time-dependent counter-rotating term $\sigma^+ a^\dagger$ gives the $e^{i2\omega t}$ term. After integrating, it results in 
        \begin{align}
            \bra{g,2}\ket{\psi^{(2)}(t)} = \frac{g_0^2\sqrt{2}it}{4\omega} + O(t^0).
            \label{eq:resonance}
        \end{align}

        The $\bra{g,2}\ket{\psi}$ state component grows linearly with time, which indicates that the integral of Eq.~(\ref{eq:second_order_int}) contains resonant terms. Moreover, this resonance increases the $\bra{g,2}\ket{\psi}$ state component, but not the amplitudes associated with the qubit excited state, $\bra{e, n}\ket{\psi}$. This result is compatible with the numerical results of Figures \ref{x-is-vt-N} and \ref{x-is-vt-Sz}, where the photon production takes place without qubit excitation. We find higher order resonances related to qubit ground state amplitudes $\bra{g,2m}\ket{\psi^{(2n)}}$ in even-order perturbation terms, whereas the odd orders do not show new resonances
        (see Appendix~\ref{app:perturbations}).
        These formulae illustrate why there is an ever-increasing photon generation without appreciable qubit excitation. In fact, there is always a pair-wise photon production, which is considered the spectral signature of the DCE \cite{nation_johansson2012, law1994}.
        
        Additionally, we consider the case where the system's frequencies are no longer resonant, but the qubit frequency is detuned by $\delta$ from the cavity mode frequency, $\Omega = \omega + \delta$. Following perturbation theory, one gets
        \begin{align}
            & \nonumber
            \psi^{(2)}(t) = \int_0^tdt'
            \int_0^{t'}dt'' \times \\
            &
            e^{-i\omega_d t'}e^{-i \delta t'}\sigma^-a^\dagger
            e^{-i\omega_d t''}e^{i(2\omega + \delta)t''}\sigma^{+}a^\dagger
            \ket{g,0}.
        \end{align}
        Here, the first integral over $t''$ results in a term $e^{i(-\omega_d + 2\omega + \delta)t'}$ that cancels the $\delta$ dependence for the second integral over $t'$. Hence, we conclude that the detuning is irrelevant in the \textit{microscopic} DCE photon generation. In contrast, the critical parameter relation is the resonance between the driving frequency and the mode frequency, $\omega_d = \omega$. In terms of the qubit velocity, the DCE is produced when the qubit's speed approaches the speed of light in the medium, $\omega_d/\omega = v/c_n \approx 1$. 

\begin{figure}
            \includegraphics[scale=0.275]{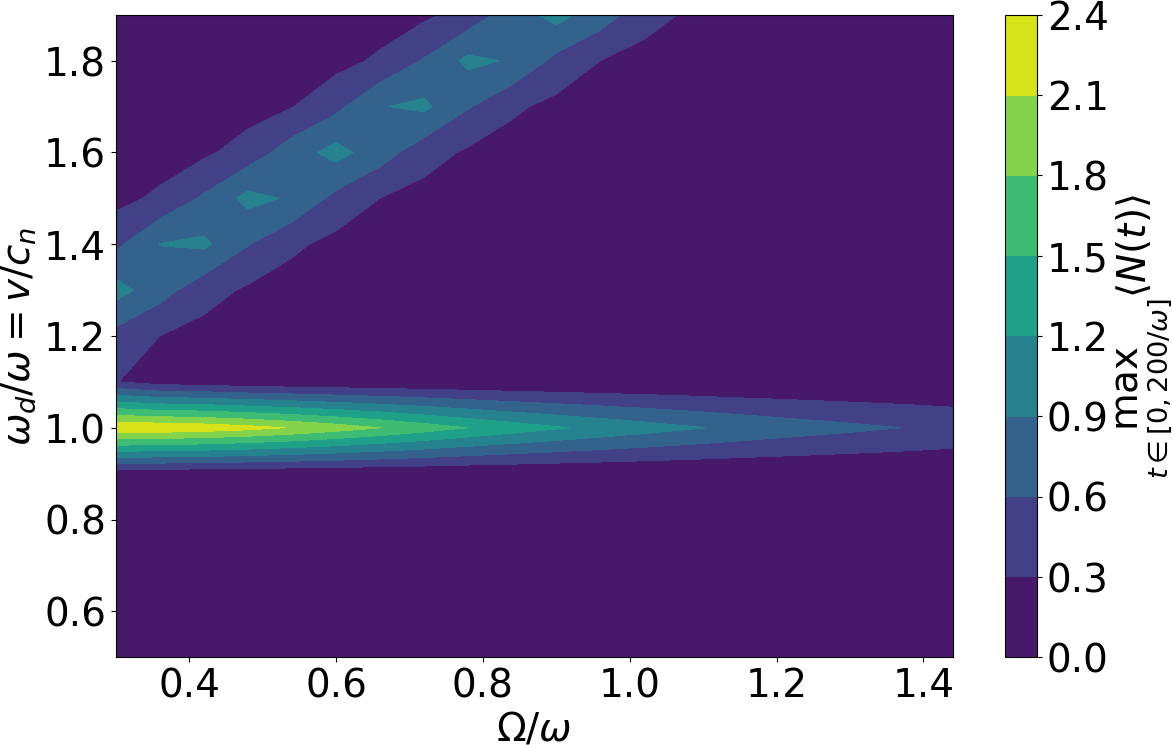}
            \caption{Maximum number of photons $\expval{N}$ in the time period $t \in [0, 200/\omega]$, that is $max_{t\in[0,200/\omega]} \expval{N(t)}$, for different values of the qubit's frequency $\Omega$ in units of the mode's frequency $\omega$ and different driving frequencies $\omega_d$ in units of the mode frequency $\omega$, or equivalently, qubit velocity $v$ in units of the speed of light in the medium $c_n$. Notice how the \textit{microscopic} DCE regime does not depend on the qubit frequency, only on its velocity which, in turn, produces the driving. On the other hand, whenever $\omega_d - \Omega = \omega$ the Unruh effect takes place and one photon and qubit excitation are produced as in a Rabi oscillation.}
            \label{wd-vs-wq-N}
        \end{figure}

        The photon generation independence on detuning is observed in Figure \ref{wd-vs-wq-N}. There, we depict the maximum number of photons over the time interval $t \in [0, 200/\omega]$ for different values of the qubit and driving frequencies, $\Omega$ and $\omega_d$, respectively. We consider a fixed mode frequency $\omega$ and a coupling intensity of $g_0 = 0.1\omega$. We note that regardless of the qubit frequency, the DCE photon generation occurs for parameter regimes near $\omega_d/\omega = v/c_n \approx 1$ as expected. Like in the original DCE, there is a transition between a non-relativistic mirror movement regime, with no photon production, and a relativistic one featuring the effect. Moreover, increasing the qubit frequency reduces the photon generation as expected from analytical calculations, since $\Omega$ appears in the denominator of the perturbative corrections. Finally, the cavity-enhanced Unruh effect appears in the superluminal regime, shown with the diagonal line $\omega_d = \Omega + \omega$ of Figure \ref{wd-vs-wq-N}. In this region of the parameter space, the photon number never exceeds one, like in a Rabi oscillation.
        See Appendix~\ref{app:numerics} for tests on the accuracy of the simulations.

        \paragraph*{Experimental possibilities and conclusions.---}
        Regarding the experimental implementation, we remark that it does not require any additional sophistication compared to the measurement of acceleration radiation or the cavity-enhanced Unruh effect, either by modulating the coupling to mimic the qubit motion \cite{felicetti_sabin2015} or by actual mechanical oscillation \cite{wang_blencowe2019, doi:10.1098/rsta.2019.0224}.
        See Appendix \ref{app:experiment} for further discussions on realistic experimental parameters that may accommodate both effects and dissipation.

        Summarizing, we have found that a discrete mirror composed of a moving qubit reproduces features of the DCE, such as photon generation from the vacuum. This photon generation takes place regardless of the qubit's internal structure and without changing its initial ground state, which supports the hypothesis that the qubit captures the essential features of a microscopic description for a moving mirror. This new effect is different from the already known cavity-enhanced Unruh effect, where the excitation of the qubit always accompanies the photon production. The \textit{microscopic} DCE explored here also differs from a more idealized proposal consisting of an atom oscillating in free space \cite{souza_impens2018}. In that case, the oscillation frequency must match the sum of the frequencies of two electromagnetic modes that, in turn, must be very small compared to the atom's internal frequency. If we translate those requirements into our system, we find the scenario of a largely detuned qubit oscillating at twice the frequency of the cavity mode, a different regime than the one found to produce the DCE in a cavity. We can relate our microscopic DCE model to other scenarios in which the RW or anti-RW approximations do not hold, joining a broad family of other settings such as the Bloch-Siegert shift \cite{forn-diaz_lisenfeld2010} or corrections on the quantum Zeno effect \cite{zheng_zhu2008}. Finally, our proposal for observing both phenomena, the cavity-enhanced Unruh effect and the microscopic DCE, could all be achieved in the same experiment, either with superconducting circuits or mechanical oscillators.
    
        A.A. and C.S have received financial support through the Postdoctoral Junior Leader Fellowship Programme from la Caixa Banking Foundation (LCF/BQ/LR18/11640005).
        L.G-Á. acknowledges support from the  Knut and Alice Wallenberg Foundation through the Wallenberg  Center for Quantum Technology (WACQT).
        E.S. acknowledges financial support from Spanish MCIU/AEI/FEDER (PGC2018-095113-B-I00), Basque Government IT986-16, projects QMiCS (820505) and Open- SuperQ (820363) of EU Flagship on Quantum Technologies, EU FET Open Grant Quromorphic, and Shanghai STCSM (Grant No. 2019SHZDZX01-ZX04).

        \appendix
        
            \section{Numerical simulation details}
            \label{app:numerics}
                We consider a system composed of a qubit of fixed frequency $\Omega$ moving within a cavity of length $L$ and coupled to its fundamental mode of frequency $\omega$ and wavenumber $k=\pi/L$. Due to said movement, the coupling oscillates in time as $g(t) = g_0\cos(kx(t))$, with $x(t)$ the trajectory of the qubit. The system is described by the Hamiltonian of Eq. (\ref{eq:H}) that we rewrite here for convenience,
                \begin{align}
                    \label{eq:Hrepeat}
                    H_{\text{total}} &= H_0 + H_{\text{int}} \nonumber \\
                    H_0 &= \frac{\Omega}{2} \sigma_z + \omega a^\dagger a \nonumber \\
                    H_{\text{int}} &= g(t)\sigma_x(a^\dagger + a).
                \end{align}
                We simulate the dynamics generated by the previous time-dependent Hamiltonian with the QuTiP library (version 4.4.1) in Python \cite{qutip}. We consider an idealized two-level qubit (that is, we neglect higher energy level excitations) coupled to a cavity fundamental mode, represented by a Hilbert space truncated to dimension $8$ that comprises the vacuum and photon number states up to $\ket{7}$.
                \begin{figure}[b!]
                    \centering
                    \includegraphics[scale=0.275]{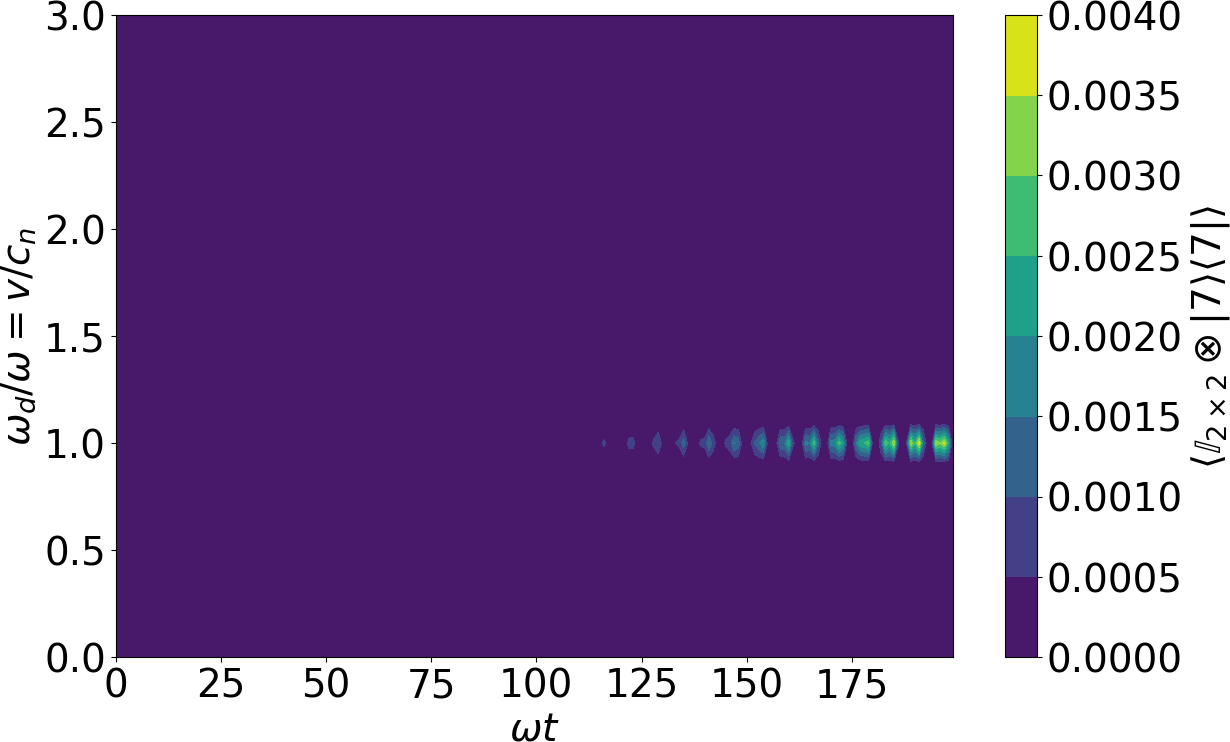}
                    \caption{Expectation value of the projector on the cutoff cavity excited state, $\mathbb{I}_{2\times2}\otimes\ket{7}\bra{7}$, as a function of time $t$ in units of the cavity frequency $\omega$, for different qubit velocities related to the driving frequency $\omega_d$, given as well in units of the cavity frequency $\omega$. We consider the same parameter domain as in Figures~\ref{x-is-vt-N} and \ref{x-is-vt-Sz}. The qubit frequency is given by $\Omega = \omega$, and the coupling intensity is $g_0 = 0.1\omega$. The qubit moves back and forth within the cavity, with constant velocity $v = L/\pi\omega_d$ in one direction, and $-v$ after bouncing in the opposite direction.}
                    \label{x-is-vt-last}
                \end{figure}
                \begin{figure}
                    \centering
                    \includegraphics[scale=0.275]{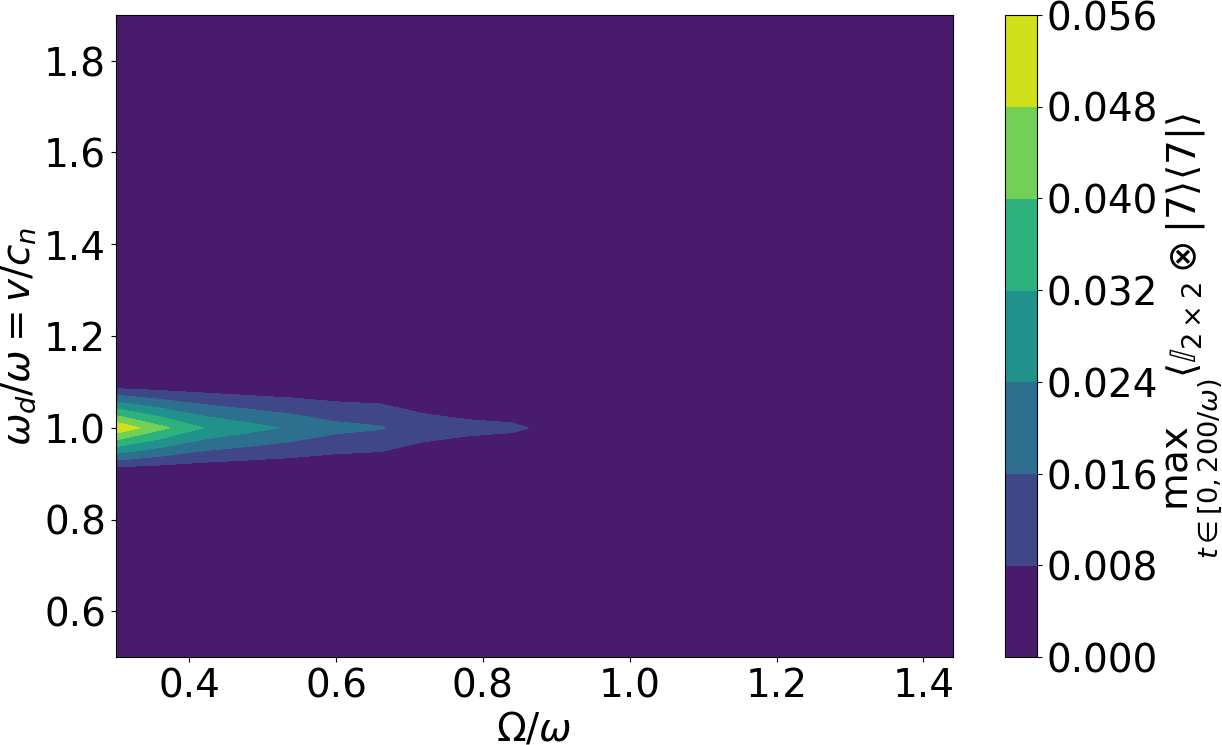}
                    \caption{Maximum value over the time period $0<t<200/\omega$ of the cutoff cavity excited state expectation value, that is $\max_{t\in[0,200/\omega)} \expval{ \mathbb{I}_{2\times2}\otimes\ket{7}\bra{7}}$, as a function of the qubit and driving frequencies, $\Omega$ and $\omega_d$ respectively, both in units of the cavity frequency $\omega$. We consider the same parameter domain as in Figure~\ref{wd-vs-wq-N}, and a coupling strength $g_0 = 0.1\omega$. The qubit moves back and forth within the cavity at a constant speed $\abs{v} = L/\pi\omega_d$.}
                    \label{wd-vs-wq-last}
                \end{figure}
                Given the nature of the dynamical Casimir effect, we expect a monotonic and unbounded parametric generation of photons \cite{uhlmann2004}. Thus, firstly, we ensure that the truncated state-space used in our calculations is large enough to describe the system's dynamics for the analyzed time interval.
                The Hamiltonian of Eq.~(\ref{eq:Hrepeat}) produces one photon per perturbation order (see details in Appendix \ref{app:perturbations}), and the vacuum state cannot evolve to a high photon number state directly. We limit the evolution time in the simulations such that the system does not reach the cutoff photon number state $\ket{7}$ from the initial low-energy states with one-photon transitions. To verify the validity of the Hilbert space truncation, we numerically confirm that the probability of measuring the cavity state $\ket{7}$ for the given time interval is negligible, as we observe in Figures \ref{x-is-vt-last} and \ref{wd-vs-wq-last}. The expectation value of the cutoff cavity state, $\expval{\mathbb{I}_{2\times 2} \otimes \ketbra{7}}$, begins acquiring significant values for $\omega_d=\omega$ and $\omega>\Omega$, precisely the parameter regimes expected to produce the DCE.
                
                Secondly, we address the numerical results congruence with the perturbative formulae in the main text. In principle, we numerically compute the dynamics for times exceeding the interval in which perturbation theory is valid. In every simulation, we use a coupling strength of $g_0=0.1\omega$ for time limits of $\omega t \approx 200$, while perturbation theory provides an accurate description of the state for $gt \approx 1$, that is, for evolution times differing by an order of magnitude $\omega t \approx 10$. We confirm the agreement between the analytic predictions and our numerical results in the regime perturbative regime $\omega t < 10$.  Moreover, we observe the monotonic unbounded nature of DCE photon generation beyond the perturbative regime for our ideal model without dissipation. Further notes on decoherence and experimental requirements can be found in Appendix \ref{app:experiment}.
                        
            \section{Complete and higher order perturbative corrections}
            \label{app:perturbations}
            In the main text, we characterize the bosonic mode population and conditions under which the dynamical Casimir effect takes place by means of typical time-dependent perturbation theory. However, given the extension of the second-order corrections in the coupling's magnitude $g_0$, we only considered those terms that become relevant to the DCE. In the following we prove that no other terms produce noticeable dynamics, even those of third order, by giving the full expressions of the corrections in the series expansion of Eq.~(\ref{series}) in the main text up to third order, for a system described by the Hamiltonian of Eq.~(\ref{eq:H}). We consider the case of resonant qubit and mode frequencies $\omega = \Omega$, and the qubit moving back and forth in the cavity with a speed $|v| = \omega L/\pi$, which leads to a DCE resonant driving $\omega_d = \omega$. Explicitly, the perturbative terms of Eq.~(\ref{series}) are given by Eqs. (\ref{psi1}-\ref{psi3})

                \begin{widetext}
                    \begin{align}
                        g_0\psi^{(1)}(t) =\import{./}{psi1.tex}
                        \label{psi1}
                    \end{align}
                    \begin{align}
                        \frac{g_0^2}{2}\psi^{(2)}(t) = \import{./}{psi2.tex}
                        \label{psi2}
                    \end{align}
                    \begin{align}
                        \frac{g_0^3}{6}\psi^{(3)}(t) = \import{./}{psi3.tex}.
                        \label{psi3}
                    \end{align}
                \end{widetext}
                
            In Eqs. (\ref{psi1}-\ref{psi3}), $\ket{0_2}$ and $\ket{1_2}$ are the qubits ground and excited state and $\ket{n}$ is a photon number state with $n$ photons. Notice that there is a resonance different to the one described in Eq. (\ref{eq:resonance}) that we have not discussed. It enlarges the projection of the state onto the ground state $\ket{0_2} \otimes \ket{0}$. The presence of this resonance does not compromise the conclusions of the main text, as it is compatible with a total state composed of a relaxed qubit and an increasing number of photons.

        \section{Experimental requirements}
            \label{app:experiment}
            A detailed experimental proposal for the realization of the DCE lies beyond the scope of this manuscript. Nevertheless, we will briefly discuss experimental parameter regimes for which an implementation of the microscopic DCE model studied in the main text may be possible. We require a tunable coupling between the qubit and the cavity of magnitude $g_0 = 0.1\omega$, that is, only one order of magnitude less than the photon frequency. Moreover, we assume that the coupling can be modulated in time with a frequency $\omega_d$ comparable to the cavity frequency $\omega$. We analyze the parameter regimes achievable with microwave-frequency superconducting circuits. Within this technology, we consider two candidates: analog simulators and FBAR-driven circuits.
            
            Firstly, we propose a modified superconducting qubit coupled to a microwave cavity. If the photons have a typical frequency of, for example, $\omega = 5$ GHz then in order to produce the microscopic DCE one would have to design a qubit with frequency preferably lower, as Figure (\ref{wd-vs-wq-N}) shows that photons production is larger in that case. To continue with the example we propose $\Omega = 2$ GHz, and coupling intensity $ g_0 = 500$ MHz \cite{srinivasan_hoffman2011}. Then, the system would have to evolve for $40$ ns, a short enough time to make dissipation irrelevant since photon lifetimes are typically in the hundreds of nanoseconds \cite{Wallraff2004}. On the other hand, if the coupling intensity is lower, for example $g_0 = 50$ MHz, the time required to produce photons increases by one order of magnitude, making dissipation relevant. That will be the case for our second experimental proposal, and so we will discuss the effects of dissipation then. For now, we discuss that the qubit will not actually move in the cavity, instead, it will \textit{simulate} its movement. As Eq.~(\ref{eq:H}) shows, the only effect movement has on the Hamiltonian is changing the value of the coupling in time. Thus one could argue that as long as an experiment manages to produce that same Hamiltonian, the same phenomena will take place, even if the qubit is static. In the latter case, the qubit could produce the time-dependent coupling if its dipolar moment changed over time. We remind the reader that this analog simulator approach was taken to observe the DCE in \cite{wilson_johansson2011,lahteenmaki_paraoanu2013}. We recall that the qubit-field interaction Hamiltonian comes from 
            \begin{align}
                 H_{\text{int}} = \mathbf{\hat{d}}\cdot\mathbf{\hat{E}}(x_{\text{qubit}}(t),t) \propto (\sigma^+ + \sigma^-)(a^\dagger + a),
            \end{align}
            where $\mathbf{\hat{d}}$ is the qubit's dipolar moment operator and $\mathbf{\hat{E}}(x,t)$ is the electric field amplitude operator throughout the cavity. On one hand, if the qubit is actually moving, its coupling will change due to the different field amplitudes it will find during its trajectory. Note that the trajectory may take the qubit over length scales comparable to the field's wavelength and the dipolar approximation will still hold, as long as the qubit charge distribution can be regarded as point-like and, when added together, neutral \cite[section $A_{IV}.1.b$]{0471845264}. Superconducting qubits in the microwave regime follow those premises: they are neutrally charged circuits several orders of magnitude smaller than microwave wavelengths \cite{Wallraff2004}. On the other hand, if the qubit is static but it can change its dipolar moment
            \begin{align}
                 H_{\text{int}} = \mathbf{\hat{d}}(t)\cdot\mathbf{\hat{E}}(x_0,t)
            \end{align}
             the same Hamiltonian can be produced with an appropriate $\mathbf{\hat{d}}(t)$. Proposals and experiments with such qubits already exist \cite{srinivasan_hoffman2011} which prove their effectiveness and feasibility of the experimental parameters mentioned before.
             
             However, a few caveats may make this experiment challenging. We have been able to pinpoint four of them, namely:
             \begin{enumerate}
                 \item Modulating coupling with no qubit frequency modulation.
                 \item Modulating longitudinal coupling with no transversal coupling.
                 \item Populating other cavity modes.
                 \item Populating the third or higher levels on the physical system that models the qubit.
             \end{enumerate}
             
             The first two points are related, so we discuss them jointly. The most common way of introducing externally controlled parameters in the system is by means of superconducting interference devices (SQUIDs), which behave as nonlinear inductors that can be tuned with the external magnetic flux that passes through them. However, in our case, the circuit must be designed such that those external parameters will modify only the dipolar moment and not the two lowest levels energy gap defining the qubit (first point). Moreover, the interaction Hamiltonian must be designed to forbid transitions between global states with the same qubit state. If the latter condition is not met, a different two-level interaction Hamiltonian should be taken into account (second point), as we explain in the following. Suppose that the circuit is described by a static Hamiltonian $H_{\text{circuit}}$ plus an interaction part of the form $H_{\text{int}} = \eta O (a^\dagger + a)$. Both operators $H_{\text{circuit}}$ and $O$ act on the degrees of freedom of the circuit, and $a$ and $a^\dagger$ act on the cavity mode state. The circuit-mode coupling $\eta$ is proportional to $g_0$. When the circuit is operated as a qubit the state has to belong to the span of the two lowest eigenvectors $H_{\text{circuit}}\ket{g} = 0$ and $H_{\text{circuit}}\ket{e} = \Omega\ket{e}$. Then one can consider a new, reduced, two-level interaction Hamiltonian $H_{\text{int},2\times 2}$ given by the matrix elements $\bra{g}H_{\text{int}}\ket{g}$, $\bra{e}H_{\text{int}}\ket{e}$ and $\bra{g}H_{\text{int}}\ket{e}$, which will produce the same dynamics as long as the circuit is operated as a qubit. By expanding the reduced interaction Hamiltonian in the Pauli basis one has
             \begin{align}
                H_{\text{int}, 2\times 2} =
                \eta (|\bra{g}O\ket{e}| \sigma_x +
                |\bra{e}O\ket{e}| \sigma_z) (a^\dagger + a),
             \end{align}
             where the energies have been rescaled so that $|\bra{g}O\ket{g}|\sigma_z$ does not appear and the Pauli basis has been rotated to conveniently remove $\sigma_y$. Then, the time dependence of the coupling comes from the time dependence of the eigenvectors $\ket{g}, \ket{e}$. In addition, it is now clear why the interaction must be engineered so that no transitions between global states with the same qubit state are allowed. If that were not the case, $\bra{e}O\ket{e}$ would not be zero and a new \textit{longitudinal} coupling would appear with operator $\sigma_z(a^\dagger + a)$.
             \begin{figure}
                 \centering
                 \includegraphics[scale=0.275]{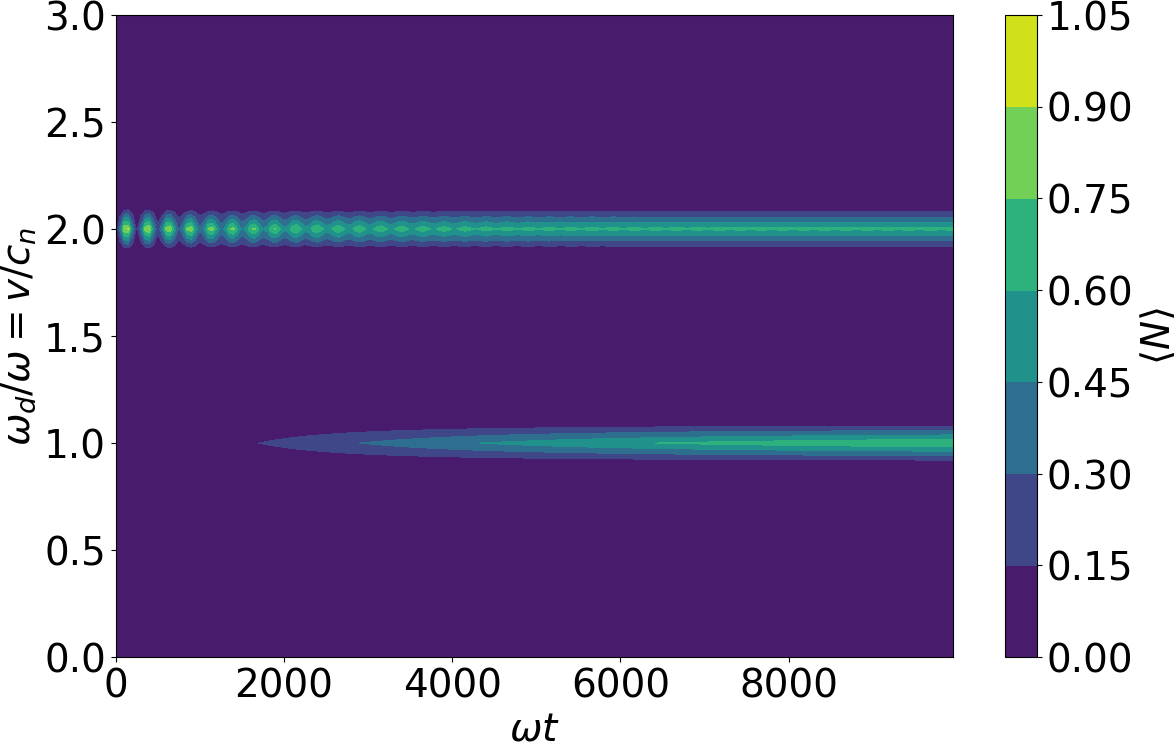}
                \caption{Number of photons $\expval{N}$ as a function of both time $t$ in units of the mode's frequency $\omega$ and driving frequency $\omega_d$ in units of the same mode's frequency $\omega$, or equivalently, qubit speed $v$ in units of the speed of light in the medium $c_n$. The qubit frequency is given by $\Omega = \omega$, and the coupling intensity is $g_0 = 0.025\omega$, to mimic the parameter regime that the experimental proposal with an actual mechanical oscillation would impose on the system. The driving frequency was produced by a qubit moving back and forth within the cavity, with constant velocity $v = L/\pi\omega_d$ in one direction, and $-v$ after bouncing in the opposite direction. We consider the collapse operators $0.025\omega a$ and $0.025\omega \sigma^-$ in the Lindblad master equation.}
                 \label{fig:decoh-N}
             \end{figure}
             \begin{figure}
                 \centering
                 \includegraphics[scale=0.275]{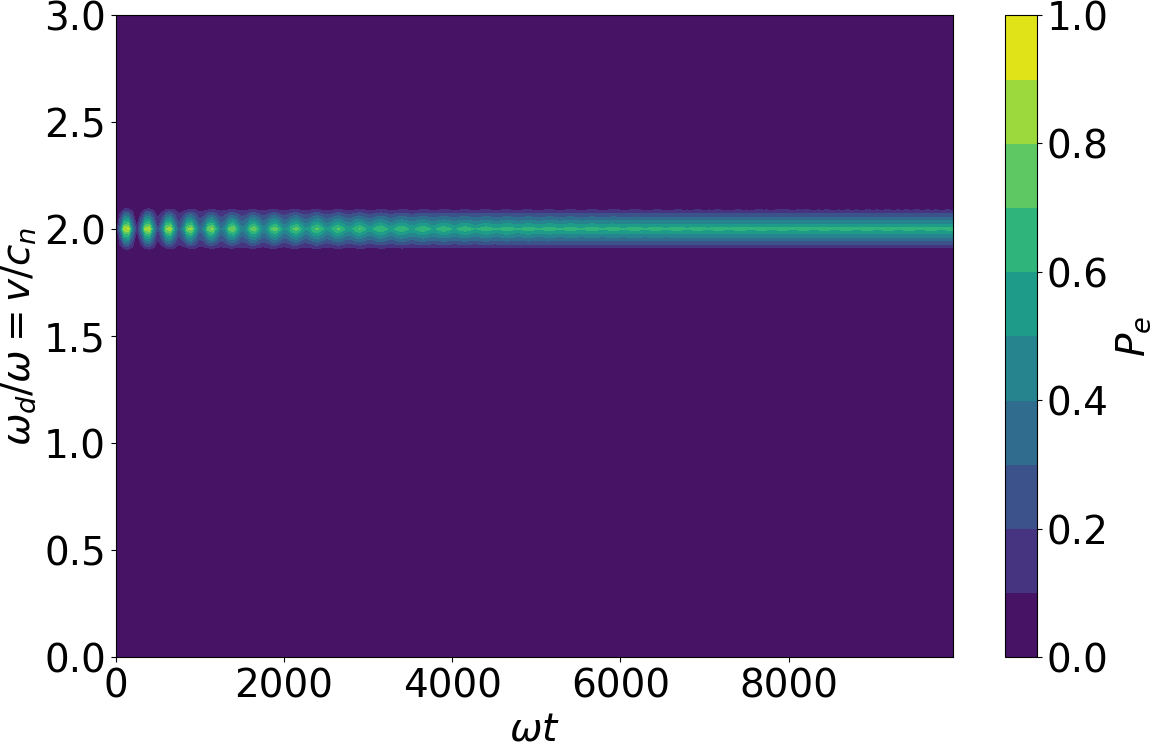}
                \caption{Population of the qubit's excited state $P_e$, that is $\expval{\sigma^+\sigma^-}$, as a function of both time $t$ in units of the mode's frequency $\omega$ and driving frequency $\omega_d$ in units of the same mode's frequency $\omega$. The qubit frequency is given by $\Omega = \omega$, and the coupling intensity is $g_0 = 0.025\omega$, to mimic the parameter regime that the experimental proposal with an actual mechanical oscillation would impose on the system. The driving frequency was produced by a qubit moving back and forth within the cavity, with constant velocity $v = L/\pi\omega_d$ in one direction, and $-v$ after bouncing in the opposite direction. We consider the collapse operators $0.025\omega a$ and $0.025\omega \sigma^-$ in the Lindblad master equation.}
                 \label{fig:decoh-Sz}
             \end{figure}
             \begin{figure}
                 \centering
                 \includegraphics[scale=0.275]{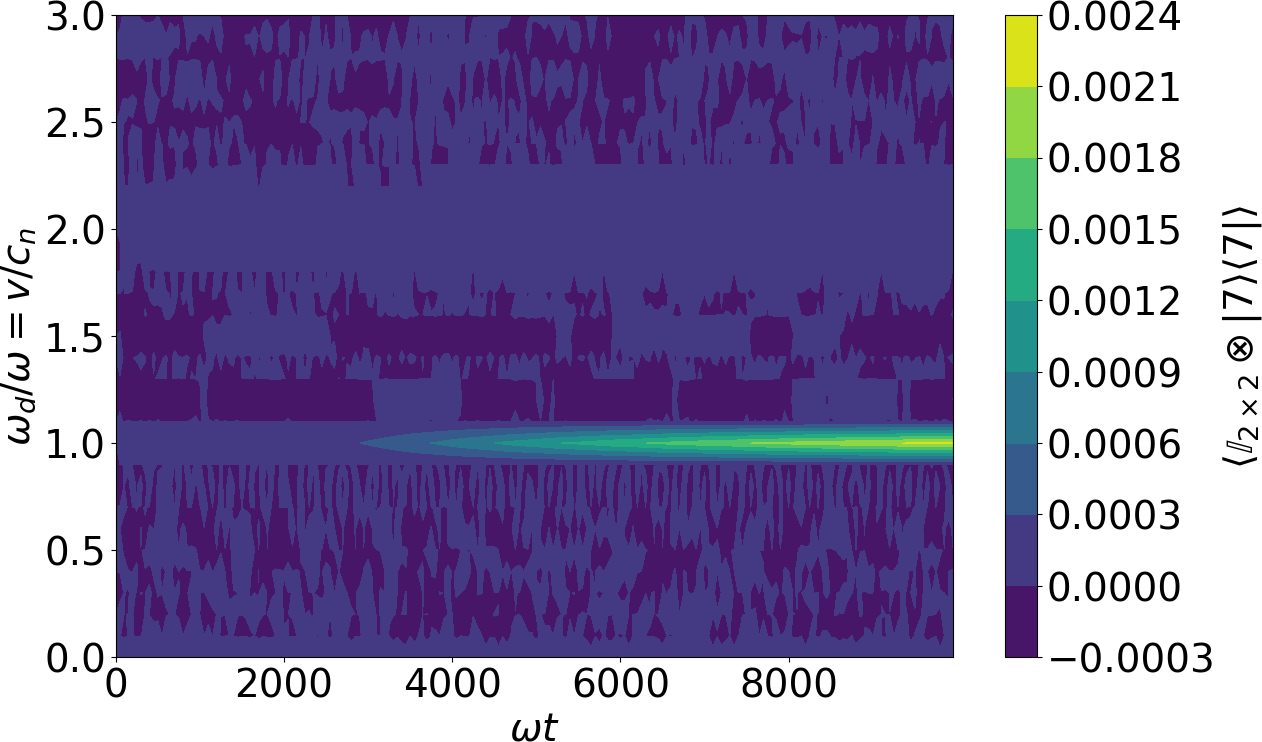}
                \caption{Expectation value of $\mathbb{I}_{2\times2}\otimes\ket{7}\bra{7}$ as a function of both time $t$ in units of the mode's frequency $\omega$ and driving frequency $\omega_d$ in units of the same mode's frequency $\omega$. The qubit frequency is given by $\Omega = \omega$, and the coupling intensity is $g_0 = 0.025\omega$, to mimic the parameter regime that the experimental proposal with an actual mechanical oscillation would impose on the system. The driving frequency was produced by a qubit moving back and forth within the cavity, with constant velocity $v = L/\pi\omega_d$ in one direction, and $-v$ after bouncing in the opposite direction. We consider the collapse operators $0.025\omega a$ and $0.025\omega \sigma^-$ in the Lindblad master equation.}
                \label{fig:decoh-last}
             \end{figure}

             For example, the tunable coupling transmon designed in \cite{srinivasan_hoffman2011} addresses satisfactorily the first point, that is, it can change the coupling intensity keeping static the qubit's frequency, but not the second. In other words, an experiment using said transmon would have to take into account a longitudinal coupling $g_z(t) \sigma_z (a^\dagger + a)$. However, it stands to reason that said transmon is a step in the right direction, and that simple modifications to its design could eliminate that piece in the Hamiltonian. As we have seen, the parameter space of the qubit comprises three parameters, frequency $\Omega$, transversal coupling $g_x$, and longitudinal $g_z$. Transmon \cite{srinivasan_hoffman2011} takes as external parameters only two magnetic fluxes, and so it can only explore a two-dimensional manifold of its three-dimensional parameter space. Thus we conclude that a similar circuit with three SQUIDs could, in principle, independently tune every parameter.

             The third point, populating other cavity modes is not a relevant issue for the microscopic DCE but it certainly is for the cavity-enhanced Unruh effect. If the mode structure is composed of equidistant modes, then producing the microscopic DCE for the fundamental mode would require a driving frequency of $\omega_d = \omega_0$. We define $\omega_0$ as the frequency of the fundamental mode, referred as $\omega$ in the main text for simplicity. Then the higher modes have frequencies $\omega_n = (n+1)\omega = (n+1)\omega_d$ which do not resonate with the driving. However, producing the cavity-enhanced Unruh effect in the fundamental mode would require a driving $\omega_d = \omega_0 + \Omega$, with $\Omega$ the frequency of the qubit. If in addition to this $\Omega \approx \omega_0$, then that same driving would produce the DCE on the next mode if frequency $\omega_1 = 2\omega \approx \Omega + \omega = \omega_d$ and both phenomena would combine in a non-trivial way. This problem can be addressed by detuning the qubit frequency $\Omega$. In fact, we find advantageous to reduce said frequency, as the velocity of the qubit required to produced the Unruh effect is $v/c_n = 1 + \Omega/\omega$. In other words, that velocity is always superluminal in the medium, but is closer to the speed of light the smaller the qubit frequency $\Omega$ is.
             
             The last caveat of the first experiment we propose is populating higher levels of the qubit system. If the qubit's complete level structure is nearly harmonic the DCE will be combined with a new resonance at second order, in which two-photon and higher-level qubit excitations take place with a magnitude comparable to the DCE. We conclude the qubit's complete level structure must be anharmonic, at least with regard to the third level, or said level will have to be taken into account.
             
             The second experiment we consider makes use of a film bulk acoustic resonator (FBAR) in order to relate the modulation of the coupling to an actual moving piece in the system. Some recent literature has considered this experiment with small variations before, we encourage the reader to read Wang \textit{et al.}\cite{wang_blencowe2019}. In that work the authors propose coupling two transmission lines by overlapping them over some of their lengths. That way, they form a capacitor which, in turn, is filled with dielectric material and cooled down to its ground mechanically oscillating level of frequency in the 1-10 GHz regime, as reported in \cite{oconnell_hofheinz2010}. The capacitor can be actuated upon by an external piezo leading to the movement of the capacitor plates. The point of \cite{wang_blencowe2019} was to interpret one of the transmission lines as a moving detector that would become excited as the other transmission line would be populated by photons by an analog Unruh effect.

             That experiment could be used for the implementation of the microscopic DCE too. In that case, one of the transmission lines should be substituted or reinterpreted as a qubit, that instead of moving back and forth in the cavity would hover up and down on top of it. The Hamiltonian of the system is very similar to Eq. \ref{eq:H}, with the difference being a  coupling directly proportional to the qubit position instead of the cosine of its position. That way the trajectory would not be one with constant velocity, but a cosine with frequency $\omega_d = \omega$. Note that the value of $\omega_d$ falls right into the microwave regime, and so cavity modes and qubit frequencies $\omega$ and $\Omega$ can be engineered in the context of microwave superconducting circuits to match it. The coupling intensity $g_0$ is connected to the lengths of the parallel strips of the FBAR by a non-trivial integral formula \cite{wang_blencowe2019}, but typical values of tens of $\mu m$ lead to couplings of $g_0 = 0.01 - 0.05 \omega$, an order of magnitude weaker than the one used in the main text. Then one must consider the evolution of the system for longer times in order to produce a measurable amount of photons. In that case, decoherence will have time to become relevant and reduce the number of photons, which raises the question of whether the DCE will be observable or not. Figures (\ref{fig:decoh-N} - \ref{fig:decoh-last}) show that the DCE photon production could be observed for the same simplified system of previous simulations evolving under a Lindbladian composed of Hamiltonian in Eq.~(\ref{eq:H}) with resonant mode's, qubit's and driving frequencies $\omega = \Omega = \omega_d$ with weak coupling $g_0 = 0.025\omega$ plus the collapse operators $0.025\omega\text{ }a$ and $0.025\omega\text{ }\sigma^-$, with $a$ the photon annihilation operator and $\sigma^-$ the qubit relaxation operator. Notice that a dissipation as intense as the coupling puts the system in a parameter regime between the strong and weak coupling regime. Such dissipation would be an overestimation, as the quantum technologies we consider have been designed to operate in the strong coupling regime since 2004 \cite{Wallraff2004}.

        \bibliography{paper.bib}{}
\end{document}

%% file: psi1.tex
\left(- \frac{2 g_{0}}{3 \omega} + \frac{g_{0} e^{3 i \omega t}}{6 \omega} + \frac{g_{0} e^{i \omega t}}{2 \omega}\right) {\ket{1_2}}\otimes {\ket{1}}

%% file: psi2.tex
\left(- \frac{13 g_{0}^{2}}{72 \omega^{2}} - \frac{g_{0}^{2} e^{- 2 i \omega t}}{16 \omega^{2}} + \frac{g_{0}^{2} e^{2 i \omega t}}{48 \omega^{2}} + \frac{g_{0}^{2} e^{- 3 i \omega t}}{18 \omega^{2}} + \frac{g_{0}^{2} e^{- i \omega t}}{6 \omega^{2}} + \frac{i g_{0}^{2} t}{6 \omega}\right) {\ket{0_2}}\otimes {\ket{0}} \nonumber \\
+ \left(- \frac{\sqrt{2} g_{0}^{2} e^{i \omega t}}{6 \omega^{2}} - \frac{3 \sqrt{2} g_{0}^{2}}{32 \omega^{2}} + \frac{\sqrt{2} g_{0}^{2} e^{4 i \omega t}}{96 \omega^{2}} + \frac{\sqrt{2} g_{0}^{2} e^{2 i \omega t}}{12 \omega^{2}} + \frac{\sqrt{2} g_{0}^{2} e^{- i \omega t}}{6 \omega^{2}} + \frac{\sqrt{2} i g_{0}^{2} t}{8 \omega}\right) {\ket{0_2}}\otimes {\ket{2}}

%% file: psi3.tex
&\left(- \frac{7 \sqrt{6} g_{0}^{3} e^{i \omega t}}{192 \omega^{3}} - \frac{\sqrt{6} g_{0}^{3} e^{4 i \omega t}}{144 \omega^{3}} - \frac{5 \sqrt{6} g_{0}^{3} e^{3 i \omega t}}{1728 \omega^{3}} + \frac{\sqrt{6} g_{0}^{3} e^{7 i \omega t}}{4032 \omega^{3}} + \frac{\sqrt{6} g_{0}^{3} e^{5 i \omega t}}{320 \omega^{3}} + \right. \nonumber \\
&\left. \frac{649 \sqrt{6} g_{0}^{3}}{15120 \omega^{3}} + \frac{\sqrt{6} i g_{0}^{3} t e^{3 i \omega t}}{144 \omega^{2}} + \frac{\sqrt{6} i g_{0}^{3} t e^{i \omega t}}{48 \omega^{2}} + \frac{\sqrt{6} i g_{0}^{3} t}{36 \omega^{2}}\right) {\ket{1_2}}\otimes {\ket{3}} \nonumber \\
&+ \left(- \frac{49 g_{0}^{3} e^{i \omega t}}{432 \omega^{3}} - \frac{7 g_{0}^{3} e^{- 2 i \omega t}}{216 \omega^{3}} - \frac{g_{0}^{3} e^{2 i \omega t}}{72 \omega^{3}} - \frac{g_{0}^{3} e^{3 i \omega t}}{648 \omega^{3}} + \frac{g_{0}^{3} e^{5 i \omega t}}{720 \omega^{3}} \right. \nonumber \\
&\left. + \frac{259 g_{0}^{3}}{1620 \omega^{3}} - \frac{i g_{0}^{3} t e^{- i \omega t}}{24 \omega^{2}} + \frac{i g_{0}^{3} t e^{3 i \omega t}}{108 \omega^{2}} + \frac{i g_{0}^{3} t}{27 \omega^{2}} + \frac{5 i g_{0}^{3} t e^{i \omega t}}{72 \omega^{2}}\right) {\ket{1_2}}\otimes {\ket{1}}